\documentclass[english]{article}
\usepackage[T1]{fontenc}
\usepackage[latin1]{inputenc}
\usepackage{a4wide}
\usepackage{subfigure}
\usepackage{graphicx}
\usepackage{makeidx}

\makeatletter
\makeindex

\newcommand{\lyxaddress}[1]{
   \par {\raggedright #1 
   \vspace{1.4em}
   \noindent\par}
 }

\usepackage{babel}
\makeatother
\begin{document}

\title{A stochastic model of wealth distribution}

\author{Indrani Bose and Subhasis Banerjee}

\maketitle
\lyxaddress{Department of Physics, Bose Institute, 93/1, A. P. C Road, Kolkata-700009, India. indrani@bosemain.boseinst.ac.in}
\section{Introduction }

A major research focus in economics and econophysics is on the distribution
of wealth in societies at different stages of development. Wealth
includes money, material goods and assets of different kinds. Knowledge
of the monetary equivalent of the latter two components is required
in order to quantify wealth. A related and easier to measure distribution
is that of income. The major motivation of theoretical models is to
provide insight on the microscopic origins of income/wealth distributions.
Such distributions are expected to provide good fits to the empirical
data. In the context of incomes, Champernowne$^{2}$ has commented
\char`\"{}The forces determining the distribution of incomes in any
community are so varied and complex and interact and fluctuate so
continuously, that any theoretical model must either be unrealistically
simplified or hopelessly complicated.\char`\"{} The statement highlights
the desirability of finding a middle ground between the unrealistically
simple and the hopelessly complicated.

A number of distribution functions has been proposed so far to describe
income and wealth distributions. Theoretical models based on stochastic
processes, have been formulated to explain the origins of some of
the distributions$^{\textrm{1-7}}$. One proposed distribution, mention
of which is found in economic literature, is the beta distribution$^{\textrm{8-9 }}$.
In this paper, we describe a simple stochastic model of wealth distribution
and show that the beta distribution is obtained in the non-equilibrium
steady state.

\section{Stochastic model}

In the model, each economic agent (can be an individual, a family
or a company) may be in two states: inactive (E) and active (E$^{\textrm{*}}$).
We determine the probability distribution of the wealth of an agent
randomly selected from a population of agents. Let the agent possess
wealth $M$ at time $t$. Increase in the wealth of the agent can
occur in two ways: at a steady rate and at random time intervals.
In state E, the agent's wealth increases at rate $b_{m}$ and in state
E$^{\textrm{*}}$, the rate is given by $b_{m}+j_{m}$. In both E
and E$^{\textrm{*}}$, the agent's wealth decreases at rate $k_{m}M$.
The decay rate is proportional to the current wealth with $k_{m}$
being the decay rate constant. Transitions between the states E and
E$^{\textrm{*}}$occur at random time intervals. The rate of change
of wealth is governed by the equation

\begin{equation}
\frac{dM}{dt}=j_{m}z+b_{m}-k_{m}M=f(M,z)\label{1}\end{equation}
where $z=1\,(0)$ when the agent is in the state E$^{\textrm{*}}$(E).
Let $p_{j}(M,t)\,\,(j=0,1)$ be the probability density function for
wealth distribution when $z=j$. The rate of change of the probability
density is given by

\begin{equation}
\frac{\partial p_{j}(M,t)}{\partial t}=-\frac{\partial}{\partial M}[f(M,z)p_{j}(M,t)]+{\displaystyle \sum_{k\neq j}}[W_{kj}p_{k}(M,t)-W_{jk}p_{j}(M,t)]\label{2}\end{equation}
where $W_{kj}$ is the transition rate from state $k$ to state $j$.
The first term in Eq.(2) is the {}``transport'' term representing
the net flow of probability density and the second term represents
the gain/loss in the probability density due to random transitions
between the state $j$ and the other accessible state. One can define
the activation and deactivation rates, $k_{a}$ and $k_{d}$ respectively,
to be $k_{a}=W_{01}$, and $k_{d}=W_{10}.$ From Eq. (2),

\begin{equation}
\frac{\partial p_{0}}{\partial t}=-\frac{\partial}{\partial M}\{(b_{m}-k_{m}M)p_{0}\}+k_{d}p_{1}-k_{a}p_{0}\label{3}\end{equation}

\begin{equation}
\frac{\partial p_{1}}{\partial t}=-\frac{\partial}{\partial M}\{(j_{m}+b_{m}-k_{m}M)p_{1}\}+k_{a}p_{0}-k_{d}p_{1}\label{4}\end{equation}
with $p=p_{0}+p_{1}$. The minimum and the maximum amounts of wealth
possessed by the agent are given by $M_{min}=b_{m}/k_{m}$ and $M_{max}=(b_{m}+j_{m})/k_{m}$.
Define $m=M/M_{max}$, $m_{min}=M_{min}/M_{max}$, $r_{1}=k_{a}/k_{m}$
and $r_{2}=k_{d}/k_{m}$. In the steady state, $\partial p_{0}/\partial t=0$
and $\partial p_{1}/\partial t=0$. The steady state solution turns
out to be the beta distribution \begin{equation}
p(m,r_{1},r_{2})=\frac{(m-m_{min})^{r_{1}-1}(1-m)^{r_{2}-1}}{B(r_{1},r_{2})(1-m_{min})^{r_{1}+r_{2}-1}}\label{5}\end{equation}
The normalization constant $B(r_{1},r_{2})$ is

\begin{equation}
B(r_{1},r_{2})=\int_{m_{min}}^{1}\frac{(m-m_{min})^{r_{1}-1}(1-m)^{r_{2}-1}}{(1-m_{min})^{r_{1}+r_{2}-1}}\,\, dm\label{6}\end{equation}
In Eqs. (5) and (6), $r_{1}>0$, $r_{2}>0$ and $m_{min}<m<1.$ Let
$m_{min}=0$ and $r_{1}>1$ and $r_{2}>1$. In this case,

\begin{equation}
B(r_{1},r_{2})=\frac{\Gamma(r_{1})\Gamma(r_{2})}{\Gamma(r_{1}+r_{2})}\label{7}\end{equation}
is the well-known beta function. The mean wealth $m_{av}$ and the
variance $m_{var}$ are given by

\begin{equation}
m_{av}=\frac{r_{1}}{r_{1}+r_{2}},\,\, m_{var}=\frac{r_{1}r_{2}}{(r_{1}+r_{2})^{2}}\,\,\frac{1}{r_{1}+r_{2}+1}\label{8}\end{equation}
The quantities depend on the ratios $r_{1}$ and $r_{2}$ rather than
on the individual values of $k_{a}$, $k_{d}$ and $k_{m}$.

\section{Results and discussion}

Societies are traditionally divided into three classes: poor, middle
and rich. Figs. 1(a), (b), and (c) show the $p(m)$ versus $m$ distributions
in the three cases. One can obtain similar curves when $m_{min}\neq0$.
The Gini coefficient $G$, a measure of wealth inequality, is expected
to be small for each separate class. For example, $G=0.2$ in the
case of Fig. 1(a) describing wealth distribution for the poor class.
The two-parameter beta distribution is flexible and can take a variety
of shapes. The precision in fitting data is, however, limited in this
case.

\begin{flushleft}%
\begin{figure}
\subfigure[]{\includegraphics[%
  width=2in,
  height=2in]{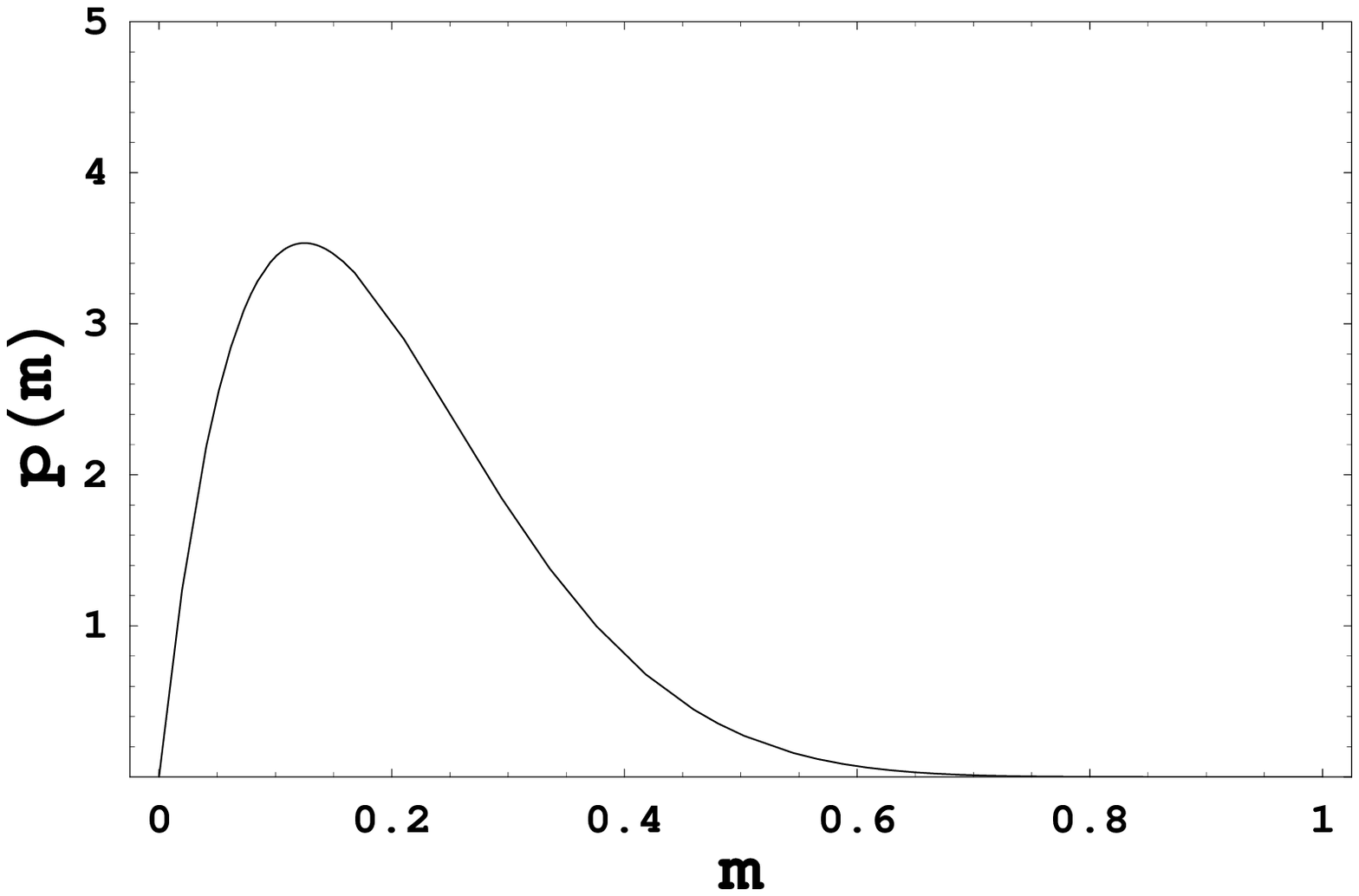}} \subfigure[]{\includegraphics[%
  width=2in,
  height=2in]{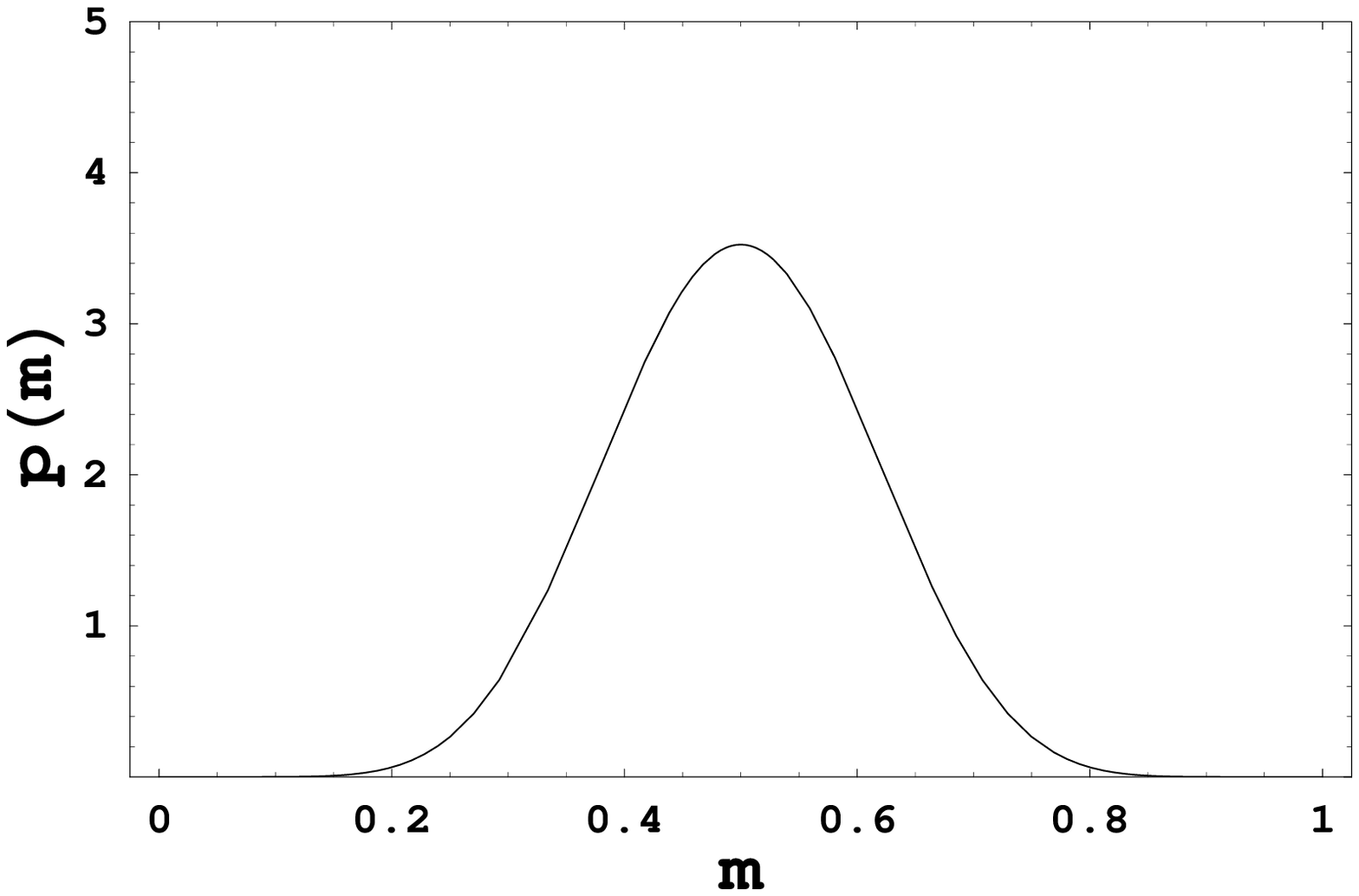}}\subfigure[]{\includegraphics[%
  width=2in,
  height=2in]{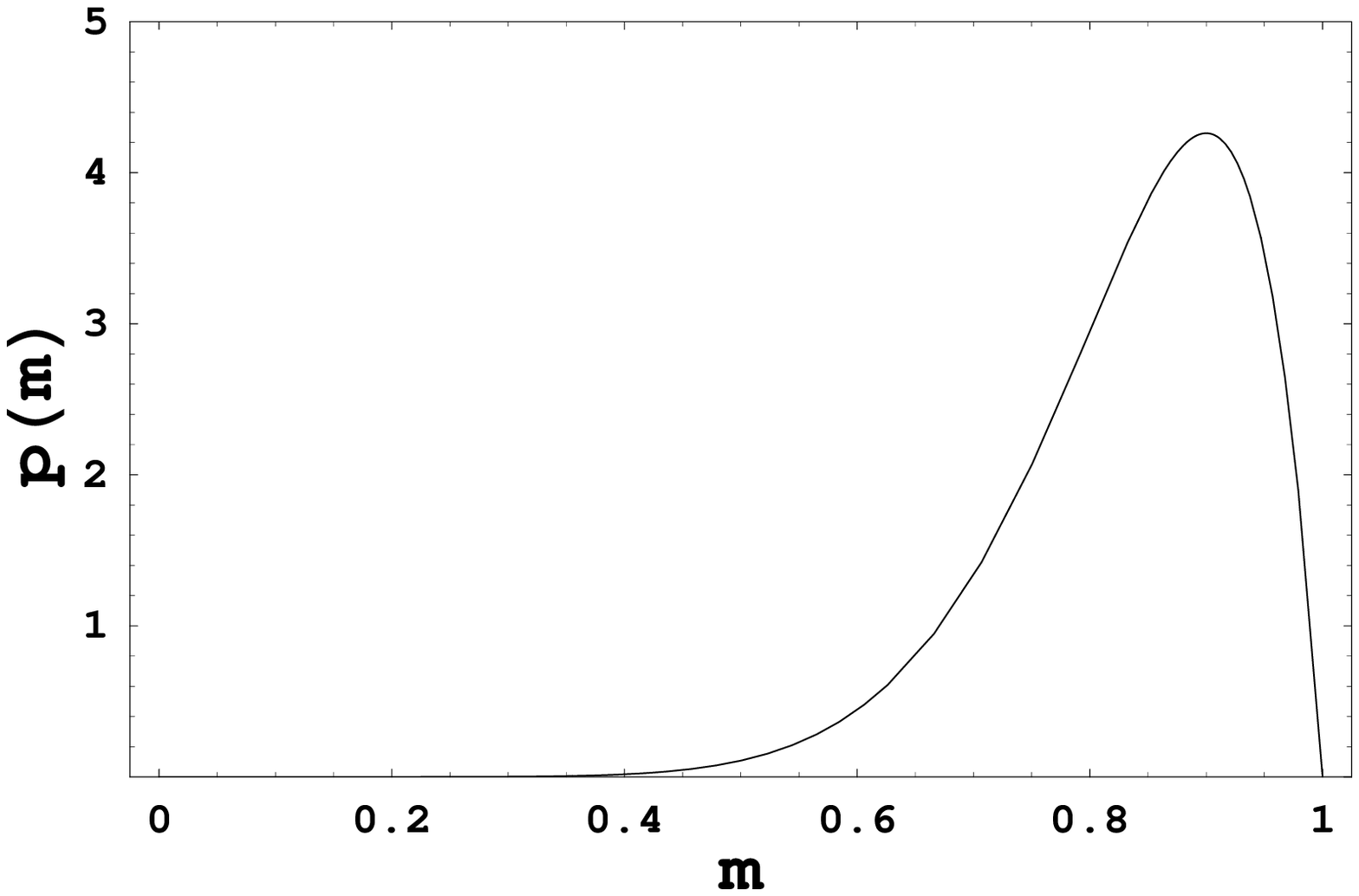}}

\caption{Probability density function $p(m)$ as a function of $m$ for (a)
$r_{1}=2,\,\, r_{2}=8,$ (b) $r_{1}=10,\,\, r_{2}=10,$ (c) $r_{1}=10,\,\, r_{2}=2$}
\label{fig:1}
\end{figure}
\end{flushleft}

\begin{flushleft}McDonald and Xu$^{10}$ have proposed a five-parameter
generalised beta distribution \begin{equation}
GB(y;\,\, a,b,c,p,q)=\frac{|a|y^{ap-1}\{1-(1-c)(\frac{y}{b})^{a}\}^{q-1}}{b^{ap}B(p,q)(1+c(\frac{y}{b})^{a})^{p+q}}\label{9}\end{equation}
where $0<y^{a}<b^{a}$ and is zero otherwise. Also, $0\leq c\leq1$
and $b,\, p,\, q>0.$ $B(p,q)$represents the normalisation constant.
The beta distribution (Eq. (5)) is a special case of GB(y; a,b,c,p,q)
with $m{}_{min}=0,$ $c=0,$ $a=1$, and $b=1$.\end{flushleft}

\includegraphics[%
  bb=6cm 6in 550bp 750bp]{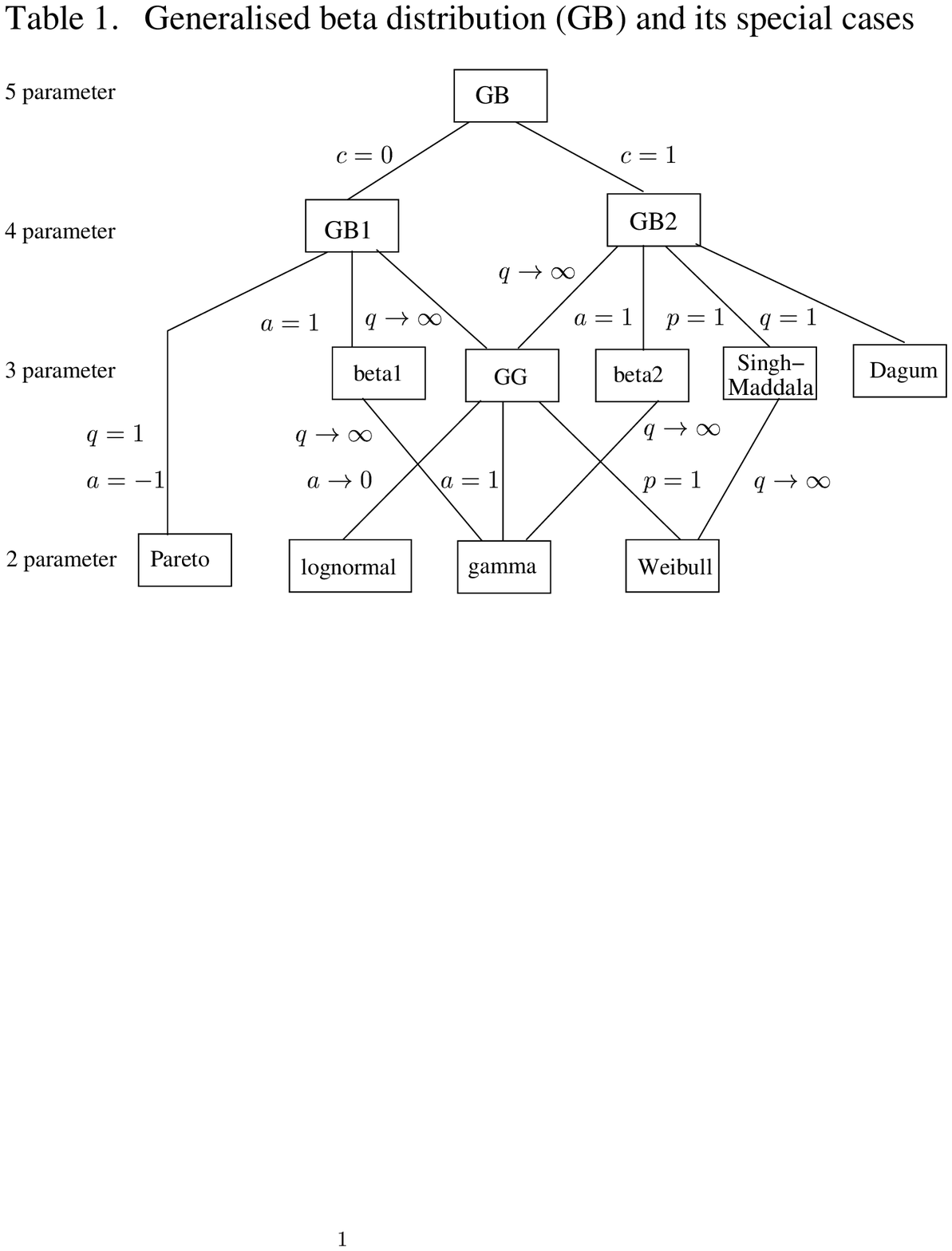}

\begin{flushleft}Many well-known distribution functions are limiting
cases of the generalised beta distribution GB. Some examples are shown
in Table 1. The beta1 distribution reduces to the beta distribution
$(m{}_{min}=0)$ with $b=1$. $GG$ refers to the generalised gamma
distribution. The special cases of $GB$$(y)$, $GB1$ and $GB2$
have been shown to outperform other distributions in providing good
quantitative fits to the income data from various countries and segments
of society. The beta distribution, considered in the paper, is a special
case of $GB1$.\end{flushleft}

In this paper, we have provided a stochastic model of wealth distribution
leading to the beta distribution in the non-equilibrium steady state.
It will be of interest to formulate stochastic models of generalised
beta distributions $GB$, $GB1$ and $GB2$. An understanding of the
microscopic origins of income/wealth distributions may provide insight
on the policies required to ensure that the benefits of economic growth
reach all sections of society.

\end{document}